\pdfoutput=1
\documentclass[12pt,a4paper]{article}
\usepackage{ifthen} % for conditional statements
\newboolean{pdflatex}
\setboolean{pdflatex}{true} % False for eps figures 

\newboolean{articletitles}
\setboolean{articletitles}{true} % False removes titles in references

\newboolean{uprightparticles}
\setboolean{uprightparticles}{false} %True for upright particle symbols

\def\paperauthors{LHCb collaboration} % Leave as is for PAPER, CONF and FIGURE
\def\paperasciititle{Measurement of the B0 -> rho0 gamma branching fraction} % Set ASCII title here !! MAKE sure it's only ASCII characters !! 
\def\papertitle{Measurement of the $\sigs$ branching fraction} % Latex formatted title
\def\paperkeywords{{High Energy Physics}, {LHCb}} % Comma separated list
\def\papercopyright{\the\year\ CERN for the benefit of the LHCb collaboration} % new since 9/Apr/2018
\def\paperlicence{CC BY 4.0 licence}
\def\paperlicenceurl{https://creativecommons.org/licenses/by/4.0/}

\newif\ifEnableSectionTOCLinks
\EnableSectionTOCLinksfalse % deactivated

\usepackage[top=1in, bottom=1.25in, left=1in, right=1in]{geometry}

\usepackage{microtype}
\usepackage{lineno}  % for line numbering during review
\usepackage{xspace} % To avoid problems with missing or double spaces after
\usepackage{caption} %these three command get the figure and table captions automatically small

\usepackage{graphicx}  % to include figures (can also use other packages)
\usepackage{color}
\usepackage{colortbl}
\graphicspath{{./figs/}} % Make Latex search fig subdir for figures
\usepackage{amsmath} % Adds a large collection of math symbols
\usepackage{amssymb}
\usepackage{amsfonts}
\usepackage{upgreek} % Adds in support for greek letters in roman typeset

\newcommand*\patchAmsMathEnvironmentForLineno[1]{%
\expandafter\let\csname old#1\expandafter\endcsname\csname #1\endcsname
\expandafter\let\csname oldend#1\expandafter\endcsname\csname
end#1\endcsname
 \renewenvironment{#1}%
   {\linenomath\csname old#1\endcsname}%
   {\csname oldend#1\endcsname\endlinenomath}%
}
\newcommand*\patchBothAmsMathEnvironmentsForLineno[1]{%
  \patchAmsMathEnvironmentForLineno{#1}%
  \patchAmsMathEnvironmentForLineno{#1*}%
}
\AtBeginDocument{%
\patchBothAmsMathEnvironmentsForLineno{equation}%
\patchBothAmsMathEnvironmentsForLineno{align}%
\patchBothAmsMathEnvironmentsForLineno{flalign}%
\patchBothAmsMathEnvironmentsForLineno{alignat}%
\patchBothAmsMathEnvironmentsForLineno{gather}%
\patchBothAmsMathEnvironmentsForLineno{multline}%
\patchBothAmsMathEnvironmentsForLineno{eqnarray}%
}

\usepackage[pdftex,
            pdfauthor={\paperauthors},
            pdftitle={\paperasciititle},
            pdfkeywords={\paperkeywords}]{hyperref}
\usepackage{hyperxmp}
\hypersetup{
    pdfcopyright={Copyright (C) \papercopyright},
    pdflicenseurl={\paperlicenceurl}
}
\usepackage[colorinlistoftodos,textsize=scriptsize]{todonotes}

\usepackage[bottom,flushmargin,hang,multiple]{footmisc}

\usepackage[all]{hypcap} % Internal hyperlinks to floats.

\usepackage{xspace}
\usepackage{upgreek}

\def\lhcb   {\mbox{LHCb}\xspace}

\def\babar  {\mbox{BaBar}\xspace}
\def\belle  {\mbox{Belle}\xspace}
\def\belletwo {\mbox{Belle~II}\xspace}

\def\ecal   {ECAL\xspace}

\def\MagUp {\mbox{\em Mag\kern -0.05em Up}\xspace}

\ifthenelse{\boolean{uprightparticles}}%
{
 
 \def\Pgamma      {\ensuremath{\upgamma}\xspace}

 \def\Peta        {\ensuremath{\upeta}\xspace}

 \def\Ppi         {\ensuremath{\uppi}\xspace}
 
 \def\Prho        {\ensuremath{\uprho}\xspace}

 \def\Pphi        {\ensuremath{\upphi}\xspace}

 \def\Pomega      {\ensuremath{\upomega}\xspace}

 \def\PDelta      {\ensuremath{\Delta}\xspace}
 \def\PXi         {\ensuremath{\Xi}\xspace}
 \def\PLambda     {\ensuremath{\Lambda}\xspace}
 \def\PSigma      {\ensuremath{\Sigma}\xspace}
 \def\POmega      {\ensuremath{\Omega}\xspace}
 \def\PUpsilon    {\ensuremath{\Upsilon}\xspace}
 \let\oldPi\Pi
 \def\PPi         {\ensuremath{\oldPi}\xspace}

 \def\PB      {\ensuremath{\mathrm{B}}\xspace}
 \def\PD      {\ensuremath{\mathrm{D}}\xspace}

 \def\PK      {\ensuremath{\mathrm{K}}\xspace}
 \def\PW      {\ensuremath{\mathrm{W}}\xspace}
 \def\Pb      {\ensuremath{\mathrm{b}}\xspace}
 \def\Pc      {\ensuremath{\mathrm{c}}\xspace}
 \def\Pd      {\ensuremath{\mathrm{d}}\xspace}

 \def\Ph      {\ensuremath{\mathrm{h}}\xspace}
 \def\Pp      {\ensuremath{\mathrm{p}}\xspace}

 \def\Ps      {\ensuremath{\mathrm{s}}\xspace}
 \def\Pt      {\ensuremath{\mathrm{t}}\xspace}

 \def\thebaroffset{0.0em}
}
{
 
 \def\Pgamma      {\ensuremath{\gamma}\xspace}

 \def\Peta        {\ensuremath{\eta}\xspace}

 \def\Ppi         {\ensuremath{\pi}\xspace}
 
 \def\Prho        {\ensuremath{\rho}\xspace}

 \def\Pphi        {\ensuremath{\phi}\xspace}

 \def\Pomega      {\ensuremath{\omega}\xspace}
 \mathchardef\PDelta="7101
 \mathchardef\PXi="7104
 \mathchardef\PLambda="7103
 \mathchardef\PSigma="7106
 \mathchardef\POmega="710A
 \mathchardef\PUpsilon="7107
 \mathchardef\PPi="7105
 \def\PB      {\ensuremath{B}\xspace}
 \def\PD      {\ensuremath{D}\xspace}

 \def\PK      {\ensuremath{K}\xspace}
 \def\PW      {\ensuremath{W}\xspace}
 \def\Pb      {\ensuremath{b}\xspace}
 \def\Pc      {\ensuremath{c}\xspace}
 \def\Pd      {\ensuremath{d}\xspace}

 \def\Ph      {\ensuremath{h}\xspace}
 \def\Pp      {\ensuremath{p}\xspace}

 \def\Ps      {\ensuremath{s}\xspace}
 \def\Pt      {\ensuremath{t}\xspace}

 \def\thebaroffset{0.18em}
}
\newcommand{\offsetoverline}[2][\thebaroffset]{\kern #1\overline{\kern -#1 #2}}%

\makeatletter
\ifcase \@ptsize \relax% 10pt
  \newcommand{\miniscule}{\@setfontsize\miniscule{4}{5}}% \tiny: 5/6
\or% 11pt
  \newcommand{\miniscule}{\@setfontsize\miniscule{5}{6}}% \tiny: 6/7
\or% 12pt
  \newcommand{\miniscule}{\@setfontsize\miniscule{5}{6}}% \tiny: 6/7
\fi
\makeatother

\DeclareRobustCommand{\optbar}[1]{\shortstack{{\miniscule (\rule[.5ex]{1.25em}{.18mm})}
  \\ [-.7ex] $#1$}}

\def\g      {{\ensuremath{\Pgamma}}\xspace}

\def\W      {{\ensuremath{\PW}}\xspace}

\def\dquark    {{\ensuremath{\Pd}}\xspace}

\def\squark    {{\ensuremath{\Ps}}\xspace}

\def\cquark    {{\ensuremath{\Pc}}\xspace}

\def\bquark    {{\ensuremath{\Pb}}\xspace}

\def\tquark    {{\ensuremath{\Pt}}\xspace}

\def\pion   {{\ensuremath{\Ppi}}\xspace}
\def\piz    {{\ensuremath{\pion^0}}\xspace}
\def\pip    {{\ensuremath{\pion^+}}\xspace}
\def\pim    {{\ensuremath{\pion^-}}\xspace}
\def\pipm   {{\ensuremath{\pion^\pm}}\xspace}
\def\pimp   {{\ensuremath{\pion^\mp}}\xspace}

\def\rhomeson {{\ensuremath{\Prho}}\xspace}
\def\rhoz     {{\ensuremath{\rhomeson^0}}\xspace}
\def\rhop     {{\ensuremath{\rhomeson^+}}\xspace}

\def\rhopm    {{\ensuremath{\rhomeson^\pm}}\xspace}

\def\kaon    {{\ensuremath{\PK}}\xspace}
\def\KorKbar {\kern \thebaroffset\optbar{\kern -\thebaroffset \PK}{}\xspace}

\def\Kp      {{\ensuremath{\kaon^+}}\xspace}
\def\Km      {{\ensuremath{\kaon^-}}\xspace}

\def\Kstarz  {{\ensuremath{\kaon^{*0}}}\xspace}

\def\Kstar   {{\ensuremath{\kaon^*}}\xspace}

\def\Kstarpm {{\ensuremath{\kaon^{*\pm}}}\xspace}

\newcommand{\phiz}{\ensuremath{\Pphi}\xspace}
\newcommand{\omegaz}{\ensuremath{\Pomega}\xspace}

\def\D       {{\ensuremath{\PD}}\xspace}

\def\DorDbar {\kern \thebaroffset\optbar{\kern -\thebaroffset \PD}\xspace}
\def\Dz      {{\ensuremath{\D^0}}\xspace}

\def\Dp      {{\ensuremath{\D^+}}\xspace}
\def\Dm      {{\ensuremath{\D^-}}\xspace}
\def\Dpm     {{\ensuremath{\D^\pm}}\xspace}

\def\DpDm    {\ensuremath{\Dp {\kern -0.16em \Dm}}\xspace}

\def\Dstarp  {{\ensuremath{\D^{*+}}}\xspace}

\def\B       {{\ensuremath{\PB}}\xspace}
\def\Bbar    {{\ensuremath{\offsetoverline{\PB}}}\xspace}

\def\BorBbar {\kern \thebaroffset\optbar{\kern -\thebaroffset \PB}\xspace}
\def\Bz      {{\ensuremath{\B^0}}\xspace}

\def\Bd      {{\ensuremath{\B^0}}\xspace}

\def\BdorBdbar {\kern \thebaroffset\optbar{\kern -\thebaroffset \Bd}\xspace}
\def\Bu      {{\ensuremath{\B^+}}\xspace}

\def\Bs      {{\ensuremath{\B^0_\squark}}\xspace}
\def\Bsb     {{\ensuremath{\Bbar{}^0_\squark}}\xspace}
\def\BsorBsbar {\kern \thebaroffset\optbar{\kern -\thebaroffset \Bs}\xspace}

\def\Y#1S{\ensuremath{\PUpsilon{(#1S)}}\xspace}

\def\proton      {{\ensuremath{\Pp}}\xspace}

\def\Lz          {{\ensuremath{\PLambda}}\xspace}

\def\LorLbar     {\kern \thebaroffset\optbar{\kern -\thebaroffset \PLambda}\xspace}

\def\Lb           {{\ensuremath{\Lz^0_\bquark}}\xspace}

\def\BF         {{\ensuremath{\mathcal{B}}}\xspace}

\newcommand{\decay}[2]{\mbox{\ensuremath{#1\!\to #2}}\xspace}

\def\to                 {\ensuremath{\rightarrow}\xspace}

\def\Vtd  {{\ensuremath{V_{\tquark\dquark}^{\phantom{\ast}}}}\xspace}

\def\Vts  {{\ensuremath{V_{\tquark\squark}^{\phantom{\ast}}}}\xspace}

\def\AT#1     {\ensuremath{A_{\mathrm{T}}^{#1}}\xspace}           % 2
\def\btosgam  {\decay{\bquark}{\squark \g}}
\def\btodgam  {\decay{\bquark}{\dquark \g}}

\def\C#1      {\ensuremath{\mathcal{C}_{#1}}\xspace}                       % 9
\def\Cp#1     {\ensuremath{\mathcal{C}_{#1}^{'}}\xspace}                    % 7
\def\Ceff#1   {\ensuremath{\mathcal{C}_{#1}^{\mathrm{(eff)}}}\xspace}        % 9  
\def\Cpeff#1  {\ensuremath{\mathcal{C}_{#1}^{'\mathrm{(eff)}}}\xspace}       % 7
\def\Ope#1    {\ensuremath{\mathcal{O}_{#1}}\xspace}                       % 2
\def\Opep#1   {\ensuremath{\mathcal{O}_{#1}^{'}}\xspace}                    % 7

\newcommand{\nospaceunit}[1]{\ensuremath{\text{#1}}}
\newcommand{\aunit}[1]{\ensuremath{\text{\,#1}}}
\newcommand{\tev}{\aunit{Te\kern -0.1em V}\xspace}
\newcommand{\gev}{\aunit{Ge\kern -0.1em V}\xspace}
\newcommand{\mev}{\aunit{Me\kern -0.1em V}\xspace}
\newcommand{\kev}{\aunit{ke\kern -0.1em V}\xspace}
\newcommand{\ev}{\aunit{e\kern -0.1em V}\xspace}

\newcommand{\mevc}{\ensuremath{\aunit{Me\kern -0.1em V\!/}c}\xspace}
\newcommand{\gevc}{\ensuremath{\aunit{Ge\kern -0.1em V\!/}c}\xspace}
\newcommand{\mevcc}{\ensuremath{\aunit{Me\kern -0.1em V\!/}c^2}\xspace}
\newcommand{\gevcc}{\ensuremath{\aunit{Ge\kern -0.1em V\!/}c^2}\xspace}
\def\mum  {\ensuremath{\,\upmu\nospaceunit{m}}\xspace}

\def\fb   {\ensuremath{\aunit{fb}}\xspace}
\def\invfb   {\ensuremath{\fb^{-1}}\xspace}

\newcommand{\chisq}{\ensuremath{\chi^2}\xspace}

\newcommand{\chisqvtx}{\ensuremath{\chi^2_{\text{vtx}}}\xspace}

\def\gsim{{~\raise.15em\hbox{$>$}\kern-.85em
          \lower.35em\hbox{$\sim$}~}\xspace}
\def\lsim{{~\raise.15em\hbox{$<$}\kern-.85em
          \lower.35em\hbox{$\sim$}~}\xspace}

\def\PDF {PDF\xspace}

\def\sPlot{\mbox{\em sPlot}\xspace}

\def\pt         {\ensuremath{p_{\mathrm{T}}}\xspace}

\def\ptot       {\ensuremath{p}\xspace}
\def\et         {\ensuremath{E_{\mathrm{T}}}\xspace}

\def\rad{\aunit{rad}\xspace}

\def\evtgen     {\mbox{\textsc{EvtGen}}\xspace}

\def\geant      {\mbox{\textsc{Geant4}}\xspace}

\def\photos     {\mbox{\textsc{Photos}}\xspace}

\def\pythia     {\mbox{\textsc{Pythia}}\xspace}

\def\tell1  {TELL1\xspace}
\def\ukl1   {UKL1\xspace}

\newcommand{\lhcborcid}[1]{\href{https://orcid.org/#1}{\hspace*{0.1em}\raisebox{-0.45ex}{\includegraphics[width=1em]{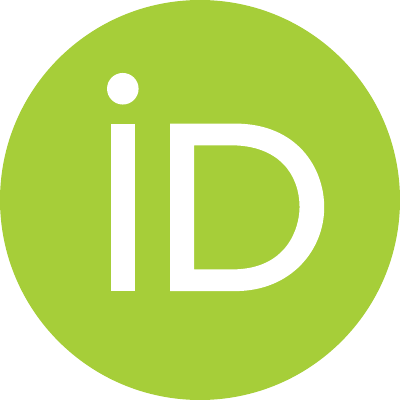}}}}

\hypersetup{
  colorlinks   = true, %Colours links instead of ugly boxes
  urlcolor     = blue, %Colour for external hyperlinks
  linkcolor    = blue, %Colour of internal links
  citecolor    = red   %Colour of citations
}

\ifEnableSectionTOCLinks
    \usepackage[explicit]{titlesec} % to change headings
    
    \let\oldcontentsline\contentsline
    \renewcommand

    \titleformat{\section}{\normalfont\Large\bf}{\hyperlink{tocsection.\thesection}{{\thesection} \parbox[t]{\dimexpr\textwidth-1pc}{#1}}}{1pc}{}

    \titleformat{\subsection}{\normalfont\bf}{\hyperlink{tocsubsection.\thesubsection}{{\thesubsection} \parbox[t]{\dimexpr\textwidth-1pc}{#1}}}{1pc}{}

    \titleformat{name=\section,numberless}[display]{}{}{0pt}{\normalfont\Huge\bfseries #1}
\fi

\usepackage{cite} % Allows for ranges in citations
\usepackage{mciteplus}
\usepackage{longtable} % only for template; not usually to be used in PAPERs

\def\perc{\ensuremath{\%}\xspace}
\def\sig {\decay{\Bd}{\rhoz(\pip \pim) \g}}
\def\ctrlnr {\decay{\Bd}{\Kstarz(\Kp \pim) \g}}
\def\sigr{\decay{\Bd}{\rhoz \g}}
\def\ctrl {\decay{\Bd}{\Kstarz \g}}

\def\sigs{\decay{\Bd}{\rho(770)^0 \g}}
\def\ctrlh {\decay{\Bd}{\Kstar(892)^0 \g}}

\def\btokstpi {\decay{\Bz}{\Kstarpm \pimp}}
\def\btorhopi {\decay{\Bz}{\rhopm \pimp}}
\def\kstkpi {\decay{\Kstarz}{\Kp\pim}}
\def\rhopipi {\decay{\rhoz}{\pip\pim}}
\begin{document}

%%%%%%%%%%%%%%%%%%%%%%%%%
%%%%% Title     %%%%%%%%%
%%%%%%%%%%%%%%%%%%%%%%%%%
\renewcommand{\thefootnote}{\fnsymbol{footnote}}
\setcounter{footnote}{1}

% %%%%%%% CHOOSE TITLE PAGE--------
%\onecolumn
% ===============================================================================
% Purpose: LHCb-PAPER journal paper title page template
% Author: 
% Created on: 2010-09-25
% ===============================================================================

%%%%%%%%%%%%%%%%%%%%%%%%%
%%%%%  TITLE PAGE  %%%%%%
%%%%%%%%%%%%%%%%%%%%%%%%%
\begin{titlepage}
\pagenumbering{roman}

% Header ---------------------------------------------------
\vspace*{-1.5cm}
\centerline{\large EUROPEAN ORGANIZATION FOR NUCLEAR RESEARCH (CERN)}
\vspace*{1.5cm}
\noindent
\begin{tabular*}{\linewidth}{lc@{\extracolsep{\fill}}r@{\extracolsep{0pt}}}
\ifthenelse{\boolean{pdflatex}}% Logo format choice
{\vspace*{-1.5cm}\mbox{\!\!\!\includegraphics[width=.14\textwidth]{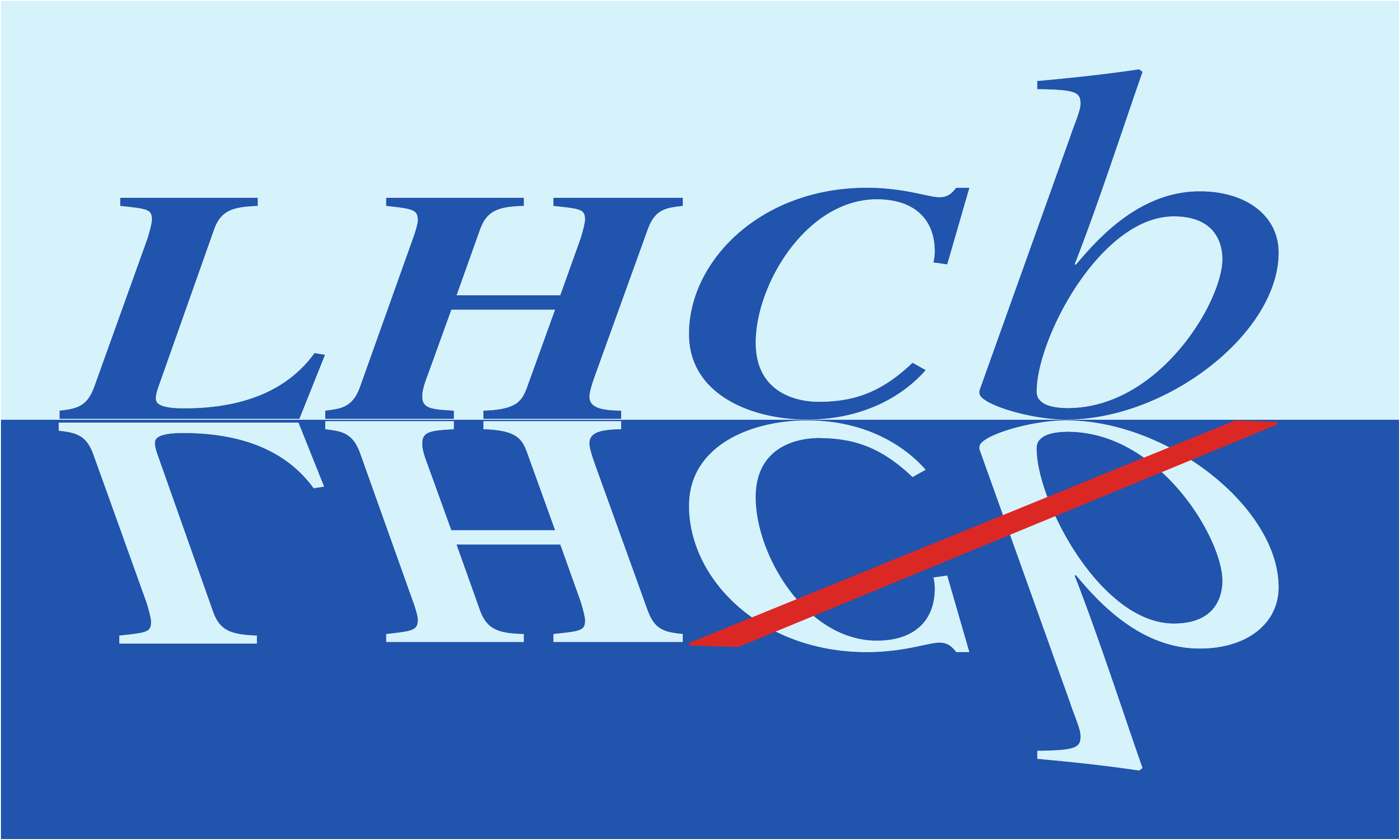}} & &}%
{\vspace*{-1.2cm}\mbox{\!\!\!\includegraphics[width=.12\textwidth]{figs/lhcb-logo.pdf}} & &}%
\\
 & & CERN-EP-2025-144 \\  % ID 
 & & LHCb-PAPER-2025-017 \\  % ID 
 & & December 19, 2025\\%19 July 2025 \\ % Date - Can also hardwire e.g.: 23 March 2010
 & & \\
% not in paper \hline
\end{tabular*}

\vspace*{3.5cm}

% Title --------------------------------------------------
{\normalfont\bfseries\boldmath\huge
\begin{center}
% DO NOT EDIT HERE. Instead edit macro in main.tex to keep metadata correct
  \papertitle 
\end{center}
}

\vspace*{2.0cm}

% Authors -------------------------------------------------
\begin{center}
%In the footnote, replace 'paper' by 'Letter' in case of submission to PRL or PLB 
% Edit macro in main.tex to keep metadata correct
\paperauthors\footnote{Authors are listed at the end of this paper.}
\end{center}

\vspace{\fill}

% Abstract -----------------------------------------------
\begin{abstract}
  \noindent
  The ratio between the branching fractions of the \sigs and \ctrlh decays is measured with proton-proton collision data collected by the \lhcb experiment at centre-of-mass energies of 7, 8, and 13\tev, corresponding to an integrated luminosity of 9\invfb. The measured value is
  \begin{equation*}
      \frac{{\BF}(\sigs)}{{\BF}(\ctrlh)}=0.0189\pm 0.0007\pm 0.0005,
  \end{equation*}
where the first uncertainty is statistical and the second systematic. The branching fraction for \sigs decays is hence obtained as
  \begin{equation*}
      {\BF}(\sigs) =(7.9\pm 0.3\pm 0.2\pm 0.2) \times 10^{-7},
  \end{equation*}
 where the last uncertainty is due to the branching fraction of the normalisation mode. This result assumes that both the $\rho(770)^0$ and $\Kstar(892)^0$ decays saturate the dihadron mass spectra considered in the analysis. It is consistent with the current world-average value and by far the most precise measurement to date.

\end{abstract}

\vspace*{2.0cm}

\begin{center}
 Published in JHEP12(2025)151  
\end{center}

\vspace{\fill}

{\footnotesize 
% Edit macro in main.tex to keep metadata correct
\centerline{\copyright~\papercopyright. \href{\paperlicenceurl}{\paperlicence}.}}
\vspace*{2mm}

\end{titlepage}

%%%%%%%%%%%%%%%%%%%%%%%%%%%%%%%%
%%%%%  EOD OF TITLE PAGE  %%%%%%
%%%%%%%%%%%%%%%%%%%%%%%%%%%%%%%%

%  empty page follows the title page ----
\newpage
\setcounter{page}{2}
\mbox{~}

%\twocolumn
% %%%%%%%%%%%%% ---------

\renewcommand{\thefootnote}{\arabic{footnote}}
\setcounter{footnote}{0}

%%%%%%%%%%%%%%%%%%%%%%%%%%%%%%%%
%%%%%  Table of Content   %%%%%%
%%%%%%%%%%%%%%%%%%%%%%%%%%%%%%%%
%%%% Uncomment if desired
%\tableofcontents

\cleardoublepage

%%%%%%%%%%%%%%%%%%%%%%%%%
%%%%% Main text %%%%%%%%%
%%%%%%%%%%%%%%%%%%%%%%%%%

\pagestyle{plain} % restore page numbers for the main text
\setcounter{page}{1}
\pagenumbering{arabic}

%% Uncomment during review phase. 
%% Comment before a final submission.
%%\linenumbers

%% This is the main body
%% It is useful to have a single file so comments are not missed in overleaf.
\section{Introduction}
\label{sec:Introduction}
In the Standard Model of particle physics~(SM), the \btodgam 
and \btosgam flavour-changing neutral transitions proceed at leading order through electroweak loop (penguin) diagrams involving virtual \W bosons and up-type quarks.
The virtual contribution of the top quark is 
dominant due to its large mass, implying these processes mostly depend on the Cabibbo--Kobayashi--Maskawa~(CKM) 
matrix elements \Vtd and \Vts, respectively. Extensions of the SM predict additional contributions that can 
significantly affect the dynamics of flavour-changing neutral transitions~\cite{Arhrib:2001gr, Ali:2002kw, Xiao:2003vq}.

The suppressed \btodgam transition is an important but relatively  unexplored probe of the SM. 
The transition was first observed in 2008 through the \sigr decay\footnote{The symbol \rhoz is used to refer to the $\rho(770)^0$ meson throughout the paper.} by both \belle~\cite{Belle2008} 
and \babar~\cite{BaBar:2008txc} experiments. A recent combination, including the \belletwo 
measurement~\cite{Belle-II:2024tru}, gives a branching fraction of 
\mbox{$\BF(\sigr)=(8.2\pm1.3)\times 10^{-7}$}~\cite{HFLAV23}.
The isospin-companion decay, $\decay{\Bu}{\rhop\g}$,\footnote{The inclusion of charge-conjugate processes is implied throughout.} is measured with a similar decay rate, 
\mbox{${\BF(\decay{\Bu}{\rhop \g})=(9.8\pm2.5)\times 10^{-7}}$}~\cite{PDG2024}. Evidence for the $\decay{\Bz}{\omegaz \g}$ decay was also reported by the \belle and \babar experiments, with a combined branching fraction of \mbox{${\BF(\decay{\Bz}{\omegaz\g})=(4.4^{+1.8}_{-1.6})\times 10^{-7}}$}~\cite{PDG2024}.

A precise measurement of the branching fraction ratio $\BF(\sigr)/\BF(\ctrl)$\footnote{The symbol \Kstarz is used to refer to the $\Kstar(892)^0$ meson throughout the paper.}
could in principle provide an independent and direct constraint on the ratio of the CKM
matrix elements $\left|{\Vtd}/{\Vts}\right|$, whose precision is
limited by the theoretical inputs in the interpretation of the oscillation frequencies of
the neutral \Bz and \Bs mesons~\cite{PDG2024}. Currently, the constraint from radiative decays suffers
from even larger theoretical uncertainties arising from the calculation of the hadronic
$\decay{\Bz}{\rhoz}$ and $\decay{\Bz}{\Kstarz}$ form factors~\cite{Ball_2006,Ball:2006eu}. Further
advancements in the form-factor calculations will be crucial to derive a competitive
constraint on the ratio $\left|{\Vtd}/{\Vts}\right|$ from the measured branching fractions
of radiative decays.

In this paper, the branching fraction of the \sigr decay is measured with respect to the  
normalisation channel, \ctrl, using the resonance decays \rhopipi and \kstkpi,
respectively. This is the first measurement
related to the \btodgam transition performed by \lhcb.
Both decay modes share a similar final-state topology allowing for the cancellation of most of the
systematic uncertainties related to the recorded luminosity, the decay reconstruction and the selection
efficiencies. The ratio of branching fractions is measured as
\begin{equation}\label{eq:RatioBF}
\frac{\BF(\sigr)}{\BF(\ctrl)} = \frac{N(\sig)}{N(\ctrlnr)}\times\frac{\varepsilon(\ctrlnr)}{\varepsilon(\sig)}\times\mathcal{R}_{\BF},
\end{equation}
where $N$ and $\varepsilon$ respectively stand for the measured yield and the overall  selection efficiency of the \sig and  
 \ctrlnr samples. The last term, \mbox{$\mathcal{R}_\BF=\BF(\kstkpi)/\BF(\rhopipi)$}, 
represents the ratio of the $\kstkpi$ and $\rhopipi$ decay rates.

The analysis uses the Run~1 (2011--2012) and Run~2 (2015--2018) \lhcb data samples of proton-proton (\proton\proton) collisions at centre-of-mass energies
of $7\tev$ (2011), $8\tev$ (2012), and $13\tev$ (Run~2), corresponding to an integrated luminosity of about 9\invfb.
To avoid experimenter’s bias, the candidates in the signal region, corresponding to an invariant mass of $m(\pip\pim\g)\in [5150,5490]\mevcc$, were not examined until the selection and analysis procedures are 
finalised. The studies were done separately per data-taking period to account for possible 
variations of the experimental conditions, and merged at the last step of the analysis.
 
\section{Detector and simulation}
\label{sec:Detector}
The \lhcb detector~\cite{LHCb-DP-2008-001,LHCb-DP-2014-002} is a single-arm forward
spectrometer covering the pseudorapidity range $2<\eta <5$,
designed for the study of particles containing \bquark or \cquark
quarks. The detector used for this analysis includes a high-precision tracking system
consisting of a silicon-strip vertex detector surrounding the \proton\proton
interaction region~\cite{LHCb-DP-2014-001}, a large-area silicon-strip detector located
upstream of a dipole magnet with a bending power of about
$4{\mathrm{\,T\,m}}$, and three stations of silicon-strip detectors and straw
drift tubes~\cite{LHCb-DP-2013-003,LHCb-DP-2017-001} placed downstream of the magnet.
The tracking system provides a measurement of the momentum, \ptot, of charged particles with
a relative uncertainty that varies from $0.5$\perc at low momentum to $1.0$\perc at $200\gevc$.
The minimum distance of a track to a primary $pp$ collision vertex (PV), the impact parameter (IP), 
is measured with a resolution of $(15+29/\pt)\mum$,
where \pt is the component of the momentum transverse to the beam, in\,\gevc.
Different types of charged hadrons are distinguished using information
from two ring-imaging Cherenkov detectors (RICH)~\cite{LHCb-DP-2012-003}. 
Photons, electrons and hadrons are identified by a calorimeter system consisting of
scintillating-pad and preshower detectors, an electromagnetic (\ecal)
and a hadronic calorimeter. Muons are identified by a
system composed of alternating layers of iron and multiwire
proportional chambers~\cite{LHCb-DP-2012-002}.

Simulated events are used to optimise the signal selection and to parameterise the reconstructed
invariant-mass distributions for the \Bz meson. Contributions from specific background channels
are also studied using dedicated simulation samples. In the simulation, $pp$ collisions are generated
using \pythia~\cite{Sjostrand:2007gs,Sjostrand:2006za} with a specific \lhcb configuration~\cite{LHCb-PROC-2010-056}. The decay chain of hadronic particles
and the final-state radiation are handled by \evtgen~\cite{Lange:2001uf} and
\photos~\cite{davidson2015photos}, respectively. The detector response
to the interacting particles is implemented in the \geant toolkit~\cite{Allison:2006ve, Agostinelli:2002hh}.

\section{Event selection}
The online event selection is performed by a trigger~\cite{LHCb-DP-2012-004, LHCb-DP-2019-001},
which consists of a hardware stage, based on information from the calorimeter and muon systems,
followed by a software stage, which applies the full event reconstruction. The hardware trigger
selects events having an \ecal cluster with an energy component transverse to the beam (\et) 
above a
threshold of $2.50\gev$ and $2.96\gev$ in Run~1 and Run~2, respectively. 
The software trigger requires two high-\pt hadrons significantly displaced from the interaction point, and one high-\et
photon~\cite{LHCb-PUB-2012-002}.  The trigger efficiency is further enhanced by about 20\perc by
applying loose track-selection requirements to events that pass a tight photon threshold of
$\et>4\gev$ at the hardware stage. For Run~2 data, a multivariate classifier based on
topological criteria complements the cut-based software trigger selection~\cite{LHCb-PROC-2015-018}.

The reconstructed \Bz candidate combines a pair of good-quality tracks~\cite{LHCb-DP-2013-002} and an energetic
photon. The two charged tracks are required to have large IP, with a
significance with respect to any PV~\cite{Kucharczyk:1756296} greater than four. Both tracks
must have a \pt larger than $500\mevc$ with at least one above $1.2\gevc$. 
To reduce the background from partially reconstructed \B decays, the \Bz decay vertex is required to be isolated
applying a lower limit
on the $\Delta\chisqvtx(\Bz)$ variable, defined as the increase in the vertex
fit \chisq when adding any additional reconstructed track.

Photon candidates are identified from \ecal clusters that cannot be geometrically associated
with any extrapolated track. The photon
four-momentum is evaluated using the dihadron vertex as the origin and the position
and energy of the associated cluster. The transverse component of the reconstructed
photon momentum is required to be larger than $3\gevc$. The \Bz candidate four-momentum
is computed by summing the four-momenta of the two hadrons and the photon. The \Bz momentum
is required to point back to the associated primary vertex and to have a transverse component
larger than $2\gevc$. The \Bz candidates are selected in the mass range $[4000,7000]\mevcc$.
The reconstructed mass of the $\pip\pim$ pair must fall within $[630,920]\mevcc$
and the $\Kp\pim$ mass within $[795.5,995.5]\mevcc$.

For the $\sig$ channel, charged hadrons are identified as pions using information provided
by the RICH system. The probability associated to the pion hypothesis must
be larger than any
other hadron hypothesis, kaon or proton, and larger than a threshold optimised to reduce the
expected contamination due to  decays of \bquark hadrons with a similar topology. The
optimisation is done by maximising the ratio \mbox{$S \,/  \sqrt{S + \sum_i B_i}$}
 where $S$ stands for the 
expected signal yield, based on the known \sigr branching fraction~\cite{PDG2024}, and $B_i$ are the expected background yields for \ctrlnr, 
$\decay{\Bs}{\phiz(\Kp\Km)\g}$ and $\decay{\Lb}{\Pp\Km\g}$ decays. This optimisation is performed for each 
year of data-taking using simulated samples and deriving particle-identification (PID) performance
from dedicated calibration-data samples. Fiducial ranges for track momentum, $p \in 
[4.5,100.0]\gevc$, and pseudorapidity, $\eta \in [1.5,4.5]$, are applied
to match the phase space covered by the data-driven calibration tool~\cite{LHCb-PUB-2016-021}. 
%%%
The charged-particle PID efficiency for the $\sig$ channel, corresponding to the probability that both pions are correctly identified, is about $70$\perc. In contrast, the PID efficiency for the \ctrlnr channel, where the kaon is misidentified as a pion, is around $0.5$\perc. 
For the \ctrlnr normalisation channel, the hadron-identification criteria were optimised in a 
previous radiative decay analysis~\cite{LHCb-PAPER-2024-002}, which uses the same procedure as that described here.

Regarding the photon identification, the main background comes from $\piz\to\g\g$ decays for which the pair of photons form a single \ecal cluster. Single photons and
neutral pions are distinguished by exploiting their cluster shape and energy distribution, using
a dedicated tool~\cite{LHCb-PUB-2015-016}. The optimisation of the photon-identification criteria is
performed with respect to the $\decay{\Bz}{\rhoz(\pip\pim)\piz}$ decay mode.
Given that its branching fraction is measured with a large relative uncertainty of $25$\perc,  the photon-identification criteria are obtained by minimising the expected relative uncertainty on the
\sig measured yield, accounting for the $\decay{\Bz}{\rhoz(\pip\pim)\piz}$ contamination and its
uncertainty. 
As for charged-particle identification, the optimisation is performed for each 
year of data-taking using simulated samples and deriving PID performance
from dedicated calibration-data samples. 
Other contributions from $\decay{\Bz}{(\pipm\piz)\Ph^\mp}$ decays, where $h$ stands for a pion or a kaon, that involve
a $(\pipm\piz)$ resonant state, including charmed \Dpm meson decays, are
vetoed by requiring that the invariant mass of the reconstructed ${\pipm\g}$ system ($m_{\pipm\g}$), where the neutral pion mass is assigned to the reconstructed photon, exceeds $2\gevcc$. This criterion, referred to as the charm veto, also suppresses contributions from charmless \btorhopi and \btokstpi decays and backgrounds from partially reconstructed \B decays involving misidentified $\piz$ mesons. 
Both the photon-identification criteria and the charm veto are consistently applied to the 
$\Kp\pim\g$ sample selection.

Background candidates resulting from combinations of unrelated hadrons and photons, denoted as combinatorial background, can be strongly suppressed by exploiting kinematic and topological variables. Boosted decision tree
(BDT) classifiers~\cite{Breiman,AdaBoost}  are trained for each year of data taking using simulated events, matched to the detector conditions, as signal proxy and data candidates selected with \mbox{$m_{\pip\pim\g}>m_{\Bz}+300\mevcc$} as
background proxy, where $m_\Bz$ is the known \Bz mass~\cite{PDG2024}. An extended dipion mass range, \mbox{$m_{\pip\pim}<1500\mevcc$}, is applied to increase
the number of $\pip\pim\g$ candidates available for the training. The input variables to the classifier are
the IP and the \pt of each hadron candidate and of the dihadron combination, as well as the momentum, pseudorapidity, flight
distance, IP significance and $\Delta\chisqvtx(\Bz)$ of the reconstructed \Bz candidate.
In Run~2, the BDT also makes use of the cone isolation variable defined as
\begin{equation}
    I_{\pt} = \frac{\pt(\Bz) - \sum \pt }{\pt(\Bz) + \sum \pt },
\end{equation}
where the sum is taken over tracks that are not part of the \Bz signal candidate but
are associated with the same PV and fall within a cone of half-angle \mbox{$\Delta {R_c}=1.7\rad$},
\mbox{$\Delta R<\Delta {R_c}$}. The track $\Delta R$ is evaluated according to \mbox{$(\Delta R)^{2}=(\Delta \theta)^{2}+(\Delta \phi)^{2}$}, where
$\Delta \theta$ and $\Delta \phi$ are the differences in the polar and azimuthal angles of the
track momentum with respect to the direction of the \Bz candidate. A two-sample cross-validation technique is used to
avoid overfitting the BDT model. A separate BDT algorithm with the same configuration and input
variables is similarly trained to select \ctrlnr candidates.

Before training the BDT, weights are applied to the simulated candidates to correct for imperfections in
the simulation of kinematic variables. They are obtained by comparing simulated
and background-subtracted data distributions for \ctrlnr, applying all the $\Kp\pim\gamma$ selection criteria
except the BDT itself, and using the \sPlot
technique~\cite{Pivk:2004ty}. A gradient-BDT 
reweighter~\cite{Rogozhnikov:2016bdp} is trained on the number of tracks per event, the  momentum and pseudorapidity of the \Bz candidate, the
transverse momenta of the photon and the dihadron candidates and, for Run~2, the cone isolation variable $I_{\pt}$.
The derived weights are applied to both the \sig and \ctrlnr simulated candidates. An overall good agreement between
data and simulation is obtained in the distributions of the variables used to train the BDT.
The effects of remaining discrepancies are considered as systematic uncertainties by using alternative weighting schemes. 

The BDT working points are optimised for each year of data-taking by maximising the ratio
\mbox{$S \,/ \sqrt{S+B}$}, where $S$ is the expected number of signal candidates estimated from
simulation and $B$ is the number of combinatorial background candidates in the signal
region estimated by extrapolating the data distribution in the upper mass sideband of
the signal peak. 

After the selection, in the $\Kp\pim\g$ sample, the expected contributions of the dominant backgrounds, relative to the \ctrlnr normalisation yield, are as follows: $0.4$\perc from 
$\decay{\Bs}{\phiz(\Kp\Km)\g}$, 
$0.5$\perc from $\decay{\Lb}{\Pp\Km\g}$ and $1.6$\perc from $\decay{\Bz}{\Kp\pim\piz}$
for which one of the particles is misidentified; $0.8$\perc from the suppressed 
$\decay{\Bsb}{\Kstarz(\Kp\pim)\g}$ mode; and 
$1.6$\perc from $\decay{\Bz}{\Kstarz(\Kp\pim)\Peta(\g\g)}$ for which one of the photons is not reconstructed. Similarly, in the
$\pip\pim\g$ sample, the expected background contributions, relative to the \sig signal yield, are as follows: $11$\perc  from $\decay{\Bz}{\Kstarz(\Kp\pim)\g}$ and $12$\perc from $\decay{\Bz}{\rhoz(\pip\pim)\piz}$ for 
which one particle is misidentified; $0.4$\perc from $\decay{\Bz}{\Kp\pim\piz}$
where both the kaon and the neutral pion are misidentified; $13$\perc from 
$\decay{\Bs}{\phiz(\pip\pim\piz)\g}$ for which the neutral pion escapes reconstruction; and
$0.9$\perc from $\decay{\Bz}{\rhoz(\pip\pim)\Peta(\g\g)}$ for which one of the photons is not
reconstructed.  For both samples, the $\decay{\Bz}{\Kp\pim\piz}$ decay is simulated using an amplitude model derived from Ref.~\cite{BaBar:2011vfx}. 

\section{Invariant-mass fit}
\label{sec:BF}
A simultaneous unbinned maximum-likelihood fit to the $\Kp\pim\g$ and $\pip\pim\g$
invariant-mass distributions is performed in the range $[5010,6410]\mevcc$ to determine the
yields of the signal and normalisation modes, combining the Run~1 and Run~2 data samples.
The invariant-mass distributions of the \sig and \ctrlnr components are modelled using a 
modified double-sided Crystal Ball (ADSCB)~\cite{Skwarnicki:1986xj} probability density function (\PDF)
with an asymmetrical
Gaussian core and tails on both sides. The mean, the width of the Gaussian core, and the 
right-tail threshold parameters are shared between the signal and the normalisation modes and are allowed
to float in the fit. 
The right-tail power parameter and the two left-tail parameters are fixed separately for each mode to the values
obtained from fits to simulation samples.

For both the $\pip\pim\g$ and $\Kp\pim\g$ mass distributions, the shapes of the 
combinatorial backgrounds are modelled using first-order polynomial functions with slopes and yields 
allowed to vary in the fit.
The backgrounds arising from partially reconstructed \B decays are modelled with an
Argus~\cite{ARGUS} function convolved with the experimental signal resolution. The slopes of the
Argus functions are fixed to their values determined from simulation. The missing-mass parameters
of the Argus functions are fixed to the mass of the \piz meson. The power parameters
and the yields are free to vary in the fit to accommodate the limited knowledge of the contributions to these missing pion components.
 
Specific background modes, listed at the end of the previous section, are included as
individual components in the fit models.
Their shapes are described with either ADSCB or Argus functions with all parameters fixed to
the values obtained from  simulation. 
Their individual contributions, to either the $\Kp\pim\g$ or $\pip\pim\g$ data
samples, are made independent of the measured \sig signal yield by defining all of them
as a fixed fraction of the fitted \ctrlnr normalisation yield.

\begin{figure}[tbph]
  \begin{center}
    \includegraphics[width=0.95\textwidth]{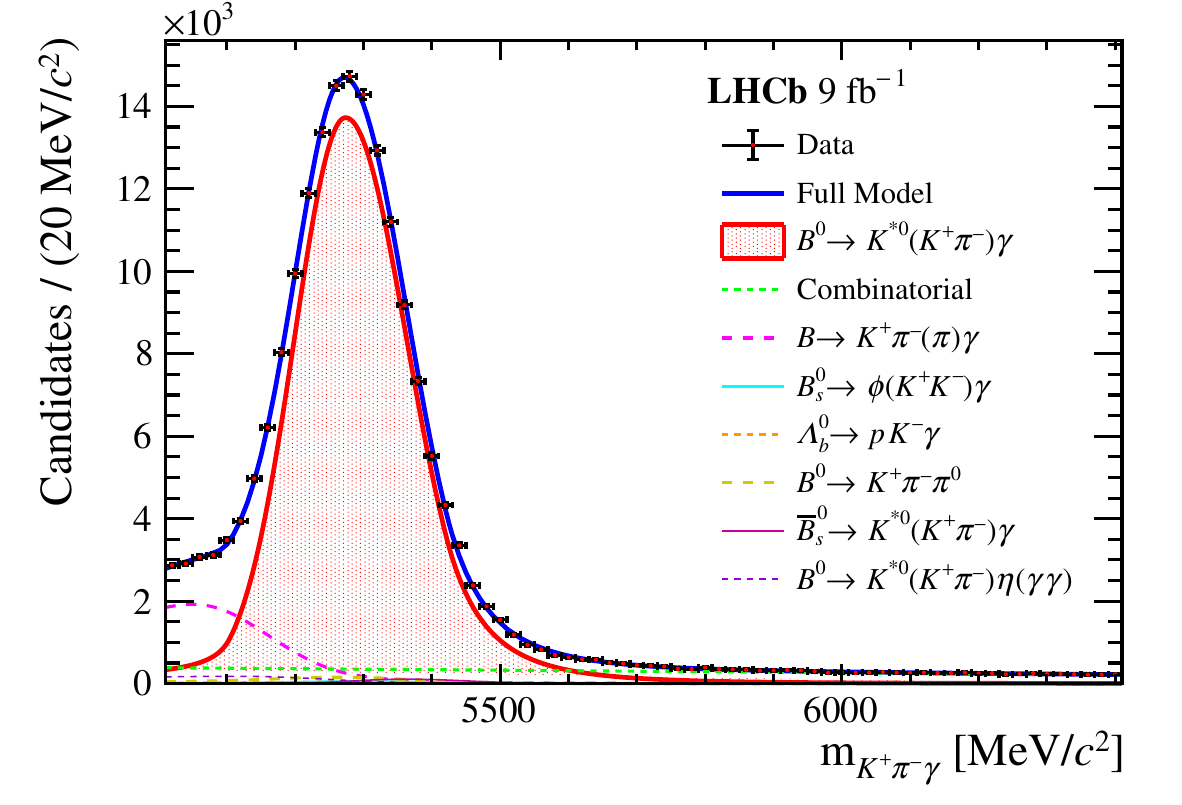}\\   
    \includegraphics[width=0.95\textwidth]{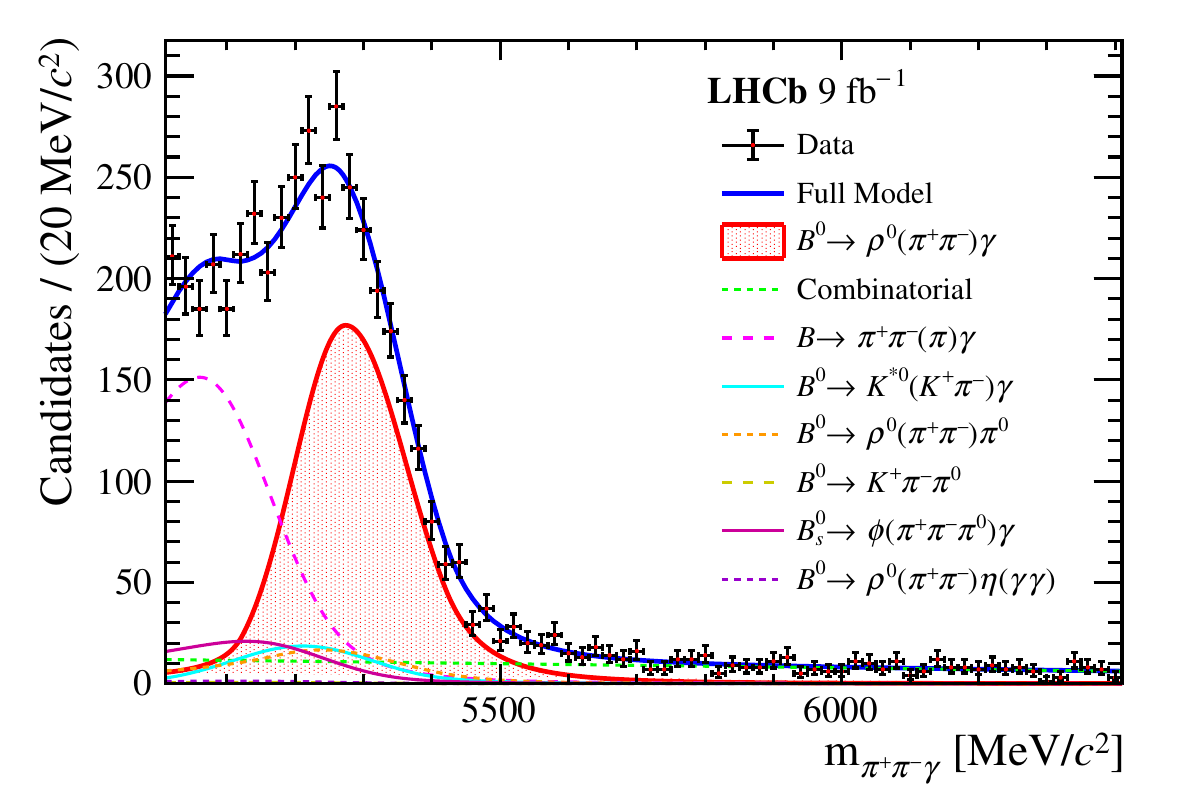}    
  \end{center}
  \caption{Invariant-mass distribution for the (top) $\Kp\pim\g$ and (bottom) $\pip\pim\g$ selected candidates 
  in the dihadron mass range $m_{\Kp\pim} \in [795.5,995.5]\mevcc$ and ${m_{\pip\pim}\in[630,920]\mevcc}$, respectively, combining Run~1 and Run~2 data samples. The fit components are also shown.}
  \label{fig:shortpipi_datafit_unblinded}
\end{figure}

The resulting invariant-mass fit is displayed in Fig.~\ref{fig:shortpipi_datafit_unblinded}.
The obtained \sig signal and \ctrlnr normalisation yields are $1\,917 \pm 71$ and $148\,160 \pm 900$, respectively, where the uncertainties are statistical only. The yield ratio is measured to be
\begin{equation*}\label{eq:Ratioyield}
\frac{N(\sig)}{N(\ctrlnr)}=0.0129\pm0.0005\pm0.0003,
\end{equation*}
where the first uncertainty is statistical and the second is the overall systematic uncertainty
resulting from the individual contributions listed in Table~\ref{t:sysyield} and summarised below.  Conservatively, individual sources are added in quadrature for the positive and negative components separately. 

\begin{table}[t]
\caption{Relative systematic uncertainties on the measured yield ratio for the different sources described in the text.}
\begin{center}\begin{tabular}{lc}
    \hline    
    Source      & Uncertainty [$\perc$] \\
    \hline
    Signal mass model          &    $(+0.5,\, -0.6)$    \\      
    Background contributions   &    $(+2.0,\, -2.2)$    \\
    Background mass models     &    $(+1.1,\, -0.8)$    \\        
     \hline
    Total systematic uncertainty   &   $(+2.3,\, -2.4)$ \\
   \hline
  \end{tabular}\end{center}
 \label{t:sysyield}
\end{table}

The uncertainty due to the fixed tail parameters in the signal model is determined using pseudoexperiments. 
The data fit is repeated 1000 times, with the value of the tail parameters sampled
according to a multidimensional Gaussian \PDF based on the covariance matrix obtained from a
simultaneous fit to the \sig and \ctrlnr simulated samples. Additionally, a potential difference in the invariant-mass resolution between signal  and normalisation modes
is investigated by introducing scale factors on the associated parameters. By adding in quadrature
the systematic uncertainties related to the signal mass model, a relative uncertainty of $\left({}^{+0.5}_{-0.6}\right)$\perc
is obtained on the yield ratio.

The specific background contributions are fixed in the simultaneous fit to the two data samples. The associated uncertainties are obtained using pseudoexperiments where each individual
contribution is varied within its uncertainty. The resulting uncertainty on the yield ratio
is found to be $\left({}^{+2.0}_{-2.2}\right)$\perc, primarily due to the limited knowledge of the branching fraction
of the $\decay{\Bz}{\rhoz(\pip\pim)\piz}$ decay mode.
Regarding the description of the background mass models, the associated uncertainties are determined
by varying the fixed parameters in the respective PDFs within their uncertainties, or by using
alternative models. Summing all the background-model sources in quadrature, an overall uncertainty of
$\left({}^{+1.1}_{-0.8}\right)$\perc on the yield ratio is obtained.

The invariant-mass fit is also repeated with the lower bound of the fitted mass range extended to
$4700\mevcc$ both for the signal and normalisation channels. Partially reconstructed 
\B decays with two or more missing pions, which populate this lower mass region are included as 
an additional component modelled with an Argus function convolved with the signal resolution. 
The missing-mass parameter of the Argus function is fixed to twice the mass of the \piz 
meson. For each of the two samples, the Argus power parameter and the associated yield
are allowed to vary freely. The obtained yield ratio is found to be compatible with the nominal
fit, and no additional uncertainty is assigned. 

\section{Efficiency ratio}
\label{sec:Efficiency}
The overall efficiency to reconstruct and select the signal and normalisation decays is factorised as 
\begin{equation}\label{eq:tot_Eff}
 \varepsilon = \varepsilon_{\text{sel}}\times \varepsilon^{\Ph^{\pm}}_{\text{PID}} \times \varepsilon^{\g}_{\text{PID}}.
\end{equation}
The term $\varepsilon_{\text{sel}}$, obtained 
from simulations, includes the geometrical acceptance of the spectrometer,
the efficiency of the trigger requirements, and the different stages of the selection
except PID criteria. 
In particular, the efficiencies related to the $\Kp\pim$ and $\pip\pim$ mass-window selections are accounted for. A relative uncertainty of $\pm0.8$\perc on the efficiency ratio, 
$\varepsilon(\ctrlnr)/\varepsilon(\sig)$, is assigned owing to the limited  size of the simulated 
samples. Alternative weighting corrections are tested for the kinematic distributions of 
the simulated samples, resulting in a systematic uncertainty of 
$\left({}^{+1.1}_{-0.2}\right)$\perc on the efficiency ratio. Differences between
kaon and pion reconstruction efficiencies lead to an additional $\pm0.3$\perc uncertainty.

The efficiencies related to the identification of the charged hadrons 
$\varepsilon^{\Ph^{\pm}}_{\text{PID}}$ are measured using high-purity calibration samples of 
$\decay{\Dstarp}{\Dz(\Km\pip)\pip}$ decays, and are evaluated as a function of the hadron momentum and 
pseudorapidity using a dedicated procedure~\cite{LHCb-PUB-2016-021}. The overall uncertainty on 
the efficiency ratio, which includes the effect of the limited size of the calibration samples, 
the correlation between the two final-state hadrons and limitations of the procedure, is estimated 
to be $\left({}^{+0.7}_{-1.3}\right)$\perc.
The photon-identification efficiency $\varepsilon^{\g}_{\text{PID}}$ is estimated from
a reference data sample using a similar method~\cite{LHCb-PUB-2015-016}. The associated
uncertainties mostly cancel in the efficiency ratio between the two radiative modes resulting in
a systematic effect estimated as $\pm0.1$\perc.

The ratio of efficiencies, obtained from a year-weighted average, is
determined to be
\begin{equation*}\label{eq:Ratioeff}
\frac{\varepsilon(\ctrlnr)}{\varepsilon(\sig)}=2.20\pm0.02\pm0.03,
\end{equation*}
where the first uncertainty is statistical and the second systematic, 
resulting from the quadratic sum of the individual contributions summarised in 
Table~\ref{t:syseff}. The value of the efficiency ratio is explained by a more stringent selection for the \sigr channel.

\begin{table}[t]
\caption{Relative systematic uncertainties on the efficiency ratio for the different sources described in the text.}
\begin{center}\begin{tabular}{lc}
    \hline    
    Source      & Uncertainty [$\perc$] \\
    \hline
    Simulated samples size    &    $(+0.8,\, -0.8)$      \\
    Kinematics corrections      &  $(+1.1,\, -0.2)$      \\
    Kaon/pion reconstruction   &  $(+0.3,\, -0.3)$       \\
    Charged PID       &    $(+0.7,\, -1.3)$              \\
    Neutral PID       &    $(+0.1,\, -0.1)$              \\
    \hline
    Total systematic uncertainty   &   $(+1.6,\, -1.6)$  \\
   \hline
  \end{tabular}\end{center}
 \label{t:syseff}
\end{table}

\section{Results}
\label{sec:conclusion}
Using the known branching fraction of \Kstarz decays into $\Kp\pim$~\cite{PDG2024}, and neglecting
any suppressed decay of the \rhoz meson, the ratio of decay rates ${\mathcal R}_b$ is taken as 
\mbox{$\BF(\kstkpi)=(66.507\pm0.014)\perc$},
which gives a ratio of branching fractions
\begin{equation*}
    \frac{\BF(\sigr)}{\BF(\ctrl)}=0.0189\pm0.0007\pm0.0005,
\end{equation*}
where the first uncertainty is statistical and the second systematic. The uncertainty on the \kstkpi branching fraction is negligible.

Combining this with the known branching
fraction for the \ctrl decay~\cite{HFLAV23}, the branching fraction
is determined to be
\begin{equation*}
    \BF(\sigr)=(7.9\pm0.3\pm0.2\pm0.2)\times 10^{-7},
\end{equation*}
where the last uncertainty is due to the branching fraction of the normalisation mode. 
The measured branching fraction is in good agreement with the current world average and is by far the most precise to date~\cite{HFLAV23}. 

The analysis of the \sig decay presented here uses candidates in the dipion invariant-mass
range $[630,920]\mevcc$ assuming that the \rhopipi decay saturates the dipion
spectrum in this range. The small contribution from $\decay{\omegaz}{\pip\pim}$ decay and any possible contribution from
wide resonances at higher mass are neglected. 
Similarly, the analysis of the \ctrlnr decay assumes that the \Kstarz decay saturates the $\Kp\pim$ invariant-mass range $[795.5,995.5]\mevcc$ considered in this study.
S-wave amplitudes are forbidden in radiative \Bz decays and nonresonant contributions with higher angular momentum are assumed to be negligible for both final states.
Detailed amplitude analyses over wider dihadron invariant-mass ranges would provide better constraints on the resonant \rhoz and \Kstarz contributions to the respective radiative decays.

% Do not include this in any draft (just for information in the template)
\section*{Acknowledgements}
%
% These Acknowledgements valid from 3-May-2019
%
\noindent We express our gratitude to our colleagues in the CERN
accelerator departments for the excellent performance of the LHC. We
thank the technical and administrative staff at the LHCb
institutes.
We acknowledge support from CERN and from the national agencies:
ARC (Australia);
CAPES, CNPq, FAPERJ and FINEP (Brazil); 
MOST and NSFC (China); 
CNRS/IN2P3 (France); 
BMBF, DFG and MPG (Germany); 
INFN (Italy); 
NWO (Netherlands); 
MNiSW and NCN (Poland); 
MCID/IFA (Romania); 
%MSHE (Russia); 
MICIU and AEI (Spain);
SNSF and SER (Switzerland); 
NASU (Ukraine); 
STFC (United Kingdom); 
DOE NP and NSF (USA).
%%%%%%%%%%%%%%%%%%%%%%%%%%%%%%%%%%%%%%%%%%%%%
We acknowledge the computing resources that are provided by ARDC (Australia), 
CBPF (Brazil),
CERN, 
IHEP and LZU (China),
IN2P3 (France), 
KIT and DESY (Germany), 
INFN (Italy), 
SURF (Netherlands),
Polish WLCG (Poland),
IFIN-HH (Romania), 
%RRCKI and Yandex LLC (Russia), 
PIC (Spain), CSCS (Switzerland), 
and GridPP (United Kingdom).
%%%%%%%%%%%%%%%%%%%%%%%%%%%%%%%%%%%%%%%%%%
We are indebted to the communities behind the multiple open-source
software packages on which we depend.
%%%%%%%%%%%%%%%%%%%%%%%%%%%%%%%%%%%%%%%%%%
Individual groups or members have received support from
%ARC and ARDC (Australia); % moved to national 16/01
Key Research Program of Frontier Sciences of CAS, CAS PIFI, CAS CCEPP, 
Fundamental Research Funds for the Central Universities,  and Sci.\ \& Tech.\ Program of Guangzhou (China);
Minciencias (Colombia);
EPLANET, Marie Sk\l{}odowska-Curie Actions, ERC and NextGenerationEU (European Union);
A*MIDEX, ANR, IPhU and Labex P2IO, and R\'{e}gion Auvergne-Rh\^{o}ne-Alpes (France);
%RFBR, RSF and Yandex LLC (Russia);
Alexander-von-Humboldt Foundation (Germany);
ICSC (Italy); 
%GVA, XuntaGal, GENCAT, Inditex, InTalent and Prog.~Atracci\'on Talento, CM (Spain);
Severo Ochoa and Mar\'ia de Maeztu Units of Excellence, GVA, XuntaGal, GENCAT, InTalent-Inditex and Prog.~Atracci\'on Talento CM (Spain);
SRC (Sweden);
the Leverhulme Trust, the Royal Society and UKRI (United Kingdom).

% Comment this in for paper drafts; do not include this in analysis note, conference and figure reports
%\input{acknowledgements}

%\input{supplementary}
%\input{appendix}

% This should be taken out in the final paper
%\input{supplementary-app}

\addcontentsline{toc}{section}{References}
%\setboolean{inbibliography}{true}
\bibliographystyle{LHCb}
\bibliography{main,standard,LHCb-PAPER,LHCb-CONF,LHCb-DP,LHCb-TDR}

@article{LHCb-DP-2019-001,
      author         = "Roel Aaij and others",
      title         = "{Design and performance of the LHCb trigger and full real-time  reconstruction in Run 2 of the LHC}",
      eprint         = "1812.10790",
      archivePrefix  = "arXiv",
      primaryClass   = "hep-ex",
      number         = "LHCb-DP-2019-001",
      year           = "2019",
      journal        = "JINST",
      volume         = "14",
      pages          = "P04013",
      doi            = "10.1088/1748-0221/14/04/P04013",
      }

@article{LHCb-DP-2017-001,
      author         = "d'Argent, Ph. and others",
      title          = "{Improved performance of the LHCb Outer Tracker in LHC
                        Run 2}",
      year           = "2017",
      eprint         = "1708.00819",
      archivePrefix  = "arXiv",
      primaryClass   = "physics.ins-det",
      report         = "LHCb-DP-2017-001",
      journal        = "JINST",
      volume         = "12",
      pages          = "P11016",
      doi            = "10.1088/1748-0221/12/11/P11016",
}

@article{LHCb-DP-2014-002,
      author         = "Aaij, R. and others",
      title          = "{LHCb detector performance}",
      collaboration  = "LHCb collaboration",
      journal        = "Int. J. Mod. Phys.",
      volume         = "A30",
      pages          = "1530022",
      doi            = "10.1142/S0217751X15300227",
      year           = "2015",
      eprint         = "1412.6352",
      archivePrefix  = "arXiv",
      primaryClass   = "hep-ex",
      report         = "LHCB-DP-2014-002, CERN-PH-EP-2014-290",
}

@article{LHCb-DP-2014-001,
      author         = "Aaij, R. and others",
      title          = "{Performance of the LHCb Vertex Locator}",
      journal        = "JINST",
      volume         = "9",
      pages          = "P09007",
      doi            = "10.1088/1748-0221/9/09/P09007",
      year           = "2014",
      eprint         = "1405.7808",
      archivePrefix  = "arXiv",
      primaryClass   = "physics.ins-det",
      report         = "LHCB-DP-2014-001",
}

@article{LHCb-DP-2013-003,
      author         = "Arink, R. and others",
      title          = "{Performance of the LHCb Outer Tracker}",
      journal        = "JINST",
      volume         = "9",
      pages          = "P01002",
      doi            = "10.1088/1748-0221/9/01/P01002",
      year           = "2014",
      eprint         = "1311.3893",
      archivePrefix  = "arXiv",
      primaryClass   = "physics.ins-det",
      report         = "LHCB-DP-2013-003",
}

@article{LHCb-DP-2013-002,
      author         = "Aaij, R. and others",
      title          = "{Measurement of the track reconstruction efficiency at LHCb}",
      collaboration  = "LHCb collaboration",
      journal        = "JINST",
      volume         = "10",
      pages          = "P02007",
      doi            = "10.1088/1748-0221/10/02/P02007",
      year           = "2015",
      eprint         = "1408.1251",
      archivePrefix  = "arXiv",
      primaryClass   = "hep-ex",
      report         = "CERN-LHCB-DP-2013-002",
}

@article{LHCb-DP-2012-004,
      author         = "Aaij, R. and others",
      title          = "{The \lhcb trigger and its performance in 2011}",
      journal        = "JINST",
      volume         = "8",
      pages          = "P04022",
      doi            = "10.1088/1748-0221/8/04/P04022",
      year           = "2013",
      eprint         = "1211.3055",
      archivePrefix  = "arXiv",
      primaryClass   = "hep-ex",
      report         = "LHCb-DP-2012-004",
}

@article{LHCb-DP-2012-003,
      author         = "Adinolfi, M. and others",
      title          = "{Performance of the \lhcb RICH detector at the LHC}",
      journal        = "Eur. Phys. J.",
      volume         = "C73",
      pages          = "2431",
      doi            = "10.1140/epjc/s10052-013-2431-9",
      year           = "2013",
      eprint         = "1211.6759",
      archivePrefix  = "arXiv",
      primaryClass   = "physics.ins-det",
      report         = "LHCb-DP-2012-003",
}

@article{LHCb-DP-2012-002,
      author       = "Alves Jr., A A and others",
      title        = "{Performance of the LHCb muon system}",
      journal      = "JINST",
      volume       = "8",
      pages        = "P02022",
      doi          = "10.1088/1748-0221/8/02/P02022",
      year         = "2013",
      eprint       = "1211.1346",
      archivePrefix= "arXiv",
      primaryClass = "physics.ins-det",
      report       = "LHCb-DP-2012-002",
}

@article{LHCb-DP-2008-001,
      author         = "Alves~Jr., A. A. and others",
      title          = "{The \lhcb detector at the LHC}",
      collaboration  = "LHCb collaboration",
      journal        = "JINST",
      volume         = "3",
      pages          = "S08005",
      doi            = "10.1088/1748-0221/3/08/S08005",
      year           = "2008",
      number         = "LHCb-DP-2008-001",
}

@article{LHCb-PAPER-2024-002,
      author         = "Aaij, R. and others",
      title          = "{Amplitude analysis of the radiative decay $B^0_s\to K^+K^-\gamma$}",
      collaboration  = "LHCb collaboration",
      report         = "{LHCb-PAPER-2024-002,  CERN-EP-2024-115}",
      eprint         = "2406.00235",
      archivePrefix  = "arXiv",
      primaryClass   = "hep-ex",
      journal           = "{JHEP}",
      volume       = "{08}",
      pages        = "{093}",
      year           = "2024",
      doi             = "10.1007/JHEP08(2024)093"
}

@article{Xiao:2003vq,
    author = "Xiao, Zhen-jun and Zhuang, Ci",
    title = "{Exclusive $B\to (K^{*}, \rho) \gamma$ decays in the general two Higgs doublet models}",
    eprint = "hep-ph/0310097",
    archivePrefix = "arXiv",
    reportNumber = "NJNU-TH-2003-17",
    doi = "10.1140/epjc/s2004-01592-0",
    journal = "Eur. Phys. J. ",
    volume = "C33",
    pages = "349--368",
    year = "2004"
}

@article{Ali:2002kw,
    author = "Ali, Ahmed and Lunghi, Enrico",
    title = "{Implications of $B \to \rho \gamma$ measurements in the standard model and supersymmetric theories}",
    eprint = "hep-ph/0206242",
    archivePrefix = "arXiv",
    reportNumber = "DESY-02-089",
    doi = "10.1140/epjc/s2002-01053-x",
    journal = "Eur. Phys. J. ",
    volume = "C26",
    pages = "195--200",
    year = "2002"
}

@article{Arhrib:2001gr,
    author = "Arhrib, Abdesslam and Chua, Chun-Khiang and Hou, Wei-Shu",
    title = "{Supersymmetric model contributions to $B^0_{d} - \bar{B}^0_d$ mixing and $B \to \pi \pi$, $\rho \gamma$ decays}",
    eprint = "hep-ph/0104122",
    archivePrefix = "arXiv",
    doi = "10.1007/s100520100758",
    journal = "Eur. Phys. J. ",
    volume = "C21",
    pages = "567--575",
    year = "2001"
}

@article{Ball:2006eu,
    author = "Ball, Patricia and Jones, Gareth W. and Zwicky, Roman",
    title = "{$B \to  V \gamma$ beyond QCD factorisation}",
    eprint = "hep-ph/0612081",
    archivePrefix = "arXiv",
    reportNumber = "IPPP-06-79, DCPT-06-158",
    doi = "10.1103/PhysRevD.75.054004",
    journal = "Phys. Rev. ",
    volume = "D75",
    pages = "054004",
    year = "2007"
}

@article{Belle2008,
       author = {{Taniguchi}, N. and others},
       collaboration = "\belle",
       title = "{Measurement of branching fractions, isospin, and CP-violating asymmetries for exclusive $b\to d \gamma$ modes}",
      journal = {Phys. Rev. Lett.},
      year = 2008,
      month = sep,
      volume = {101},
      number = {11},
      eid = {111801},
      pages = {111801},
      doi = {10.1103/PhysRevLett.101.111801},
      archivePrefix = {arXiv},
      eprint = {0804.4770},
      primaryClass = {hep-ex},
      adsurl = {https://ui.adsabs.harvard.edu/abs/2008PhRvL.101k1801T}
}

@article{BaBar:2008txc,
    author = "Aubert, Bernard and others",
    collaboration = "BaBar",
    title = "{Measurements of branching fractions for $B^{+} \to \rho^{+} \gamma$, $B^0 \to \rho^0 \gamma$, and $B^0 \to \omega \gamma$}",
    eprint = "0808.1379",
    archivePrefix = "arXiv",
    primaryClass = "hep-ex",
    reportNumber = "SLAC-PUB-13338, BABAR-PUB-08-024",
    doi = "10.1103/PhysRevD.78.112001",
    journal = "Phys. Rev. ",
    volume = "D78",
    pages = "112001",
    year = "2008"
}

@article{BaBar:2011vfx,
    author = "Lees, J. P. and others",
    collaboration = "BaBar",
    title = "{Amplitude analysis of $B^0\to K^+ \pi^- \pi^0$ and evidence of direct CP violation in $B\to K^* \pi$ decays}",
    eprint = "1105.0125",
    archivePrefix = "arXiv",
    primaryClass = "hep-ex",
    reportNumber = "SLAC-PUB-14452, BABAR-PUB-09-040",
    doi = "10.1103/PhysRevD.83.112010",
    journal = "Phys. Rev. ",
    volume = "D83",
    pages = "112010",
    year = "2011"
}

@article{Belle-II:2024tru,
  title = "{Measurement of branching fractions, $CP$ asymmetry, and isospin asymmetry for $B\rightarrow\rho\gamma$ decays using Belle and Belle II data}",
  author = "Adachi, I. and others",
  collaboration = "\belle and \belletwo collaborations",
  eprint = "2407.08984",
  archivePrefix = "arXiv",
  primaryClass = "hep-ex",
  journal = "Phys. Rev. ",
  volume = "D111",
  issue = "7",
  pages = "L071103",
  numpages = "12",
  year = "2025",
  month = "Apr",
  publisher = "American Physical Society",
  doi = "10.1103/PhysRevD.111.L071103",
  url = "https://link.aps.org/doi/10.1103/PhysRevD.111.L071103"
}

@article{Ball_2006,
    author = "Ball, Patricia and Zwicky, Roman",
    title = "{$|V_{td}/V_{ts}|$ from $B \to V\gamma$}",
    eprint = "hep-ph/0603232",
    archivePrefix = "arXiv",
    reportNumber = "IPPP-06-16, DCPT-06-32, CERN-PH-TH-2006-050",
    doi = "10.1088/1126-6708/2006/04/046",
    journal = "JHEP",
    volume = "04",
    pages = "046",
    year = "2006"    
}

@article{ARGUS,
    author = "Albrecht, H. and others",
    collaboration = "ARGUS",
    title = "{Search for hadronic $b \to u$ decays}",
    reportNumber = "DESY-90-008",
    doi = "10.1016/0370-2693(90)91293-K",
    journal = "Phys. Lett. ",
    volume = "B241",
    pages = "278--282",
    year = "1990"
}

@lhcbreport{LHCb-PUB-2015-016,
      author        = "Calvo Gomez, Miriam and Cogneras, Eric and Deschamps,
                       Olivier and Hoballah, Mostafa and Lefevre, Regis and
                       Monteil, Stephane and Puig Navarro, Albert and Rives
                       Molina, Vicente Jose",
      title         = "{A tool for $\gamma/\pi^0$ separation at high energies}",
      institution   = "CERN",
      number  = "LHCb-PUB-2015-016, CERN-LHCb-PUB-2015-016",
      address       = "Geneva",
      year          = "2015",
      url           = "https://cds.cern.ch/record/2042173",
}

@lhcbreport{LHCb-PUB-2012-002,
      author        = "Puig, Albert",
      title         = "{The HLT2 radiative topological lines}",
      number        = "LHCb-PUB-2012-002, CERN-LHCb-PUB-2012-002",
      institution   = "CERN",
      address       = "Geneva",
      year          = "2012",
      url           = "https://cds.cern.ch/record/1426295",
}

@article{PDG2024,
     author    = "Navas, S. and others",
    collaboration = "Particle Data Group",
     title     = "{\href{http://pdg.lbl.gov/}{Review of particle physics}}",
     journal   = "Phys. Rev.",
     year = {2024},
     number = {8},
     volume      = "D110",
     pages     = "030001",
     doi = "10.1103/PhysRevD.110.030001"
}

@article{LHCb-PROC-2015-018,
      author        = "Likhomanenko, T. and others",
      title         = "{LHCb topological trigger reoptimization}",
      journal       = "J. Phys. Conf. Ser.",
      volume        = "664",
      pages         = "082025",
      doi           = "10.1088/1742-6596/664/8/082025",
      month         = "Oct",
      year          = "2015",
      reportNumber  = "LHCb-PROC-2015-018",
      eprint        = "1510.00572",
      archivePrefix = "arXiv"
}

@article{LHCb-PROC-2010-056,
      author         = "Belyaev, I. and others",
      title          = "{Handling of the generation of primary events
                         in Gauss, the LHCb simulation framework}",
      journal="J. Phys. Conf. Ser.",
      volume={331},
      pages={032047},
      doi={10.1088/1742-6596/331/3/032047},
      year={2011},
}

@article{HFLAV23,
      author = "Banerjee, S. and others",
      collaboration  = "Heavy Flavor Averaging Group",
      title = "{Averages of $b$-hadron, $c$-hadron, and $\tau$-lepton properties as of 2023}",
      eprint = "2411.18639",
      archivePrefix = "arXiv",
      primaryClass = "hep-ex",
      year = "2024",
      note           = "{updated results and plots available at \href{https://hflav.web.cern.ch}{{\texttt{https://hflav.web.cern.ch}}}}",
}

@article{Sjostrand:2006za,
      author         = {Sj\"{o}strand, Torbj\"{o}rn and Mrenna, Stephen and
                       Skands, Peter"},
      title          = "{PYTHIA 6.4 physics and manual}",
      journal        = "JHEP",
      volume         = "05",
      pages          = "026",
      doi            = "10.1088/1126-6708/2006/05/026",
      year           = "2006",
      eprint         = "hep-ph/0603175",
      archivePrefix  = "arXiv",
      primaryClass   = "hep-ph",
}

@article{Sjostrand:2007gs,
      author         = {Sj\"{o}strand, Torbj\"{o}rn and Mrenna, Stephen and
                        Skands, Peter"},
      title          = "{A brief introduction to PYTHIA 8.1}",
      journal        = "Comput. Phys. Commun.",
      volume         = "178",
      pages          = "852-867",
      doi            = "10.1016/j.cpc.2008.01.036",
      year           = "2008",
      eprint         = "0710.3820",
      archivePrefix  = "arXiv",
      primaryClass   = "hep-ph",
      reportNumber   = "CERN-LCGAPP-2007-04, LU-TP-07-28,
                        FERMILAB-PUB-07-512-CD-T",
}

@Article{Agostinelli:2002hh,
     author    = "Agostinelli, S. and others",
 collaboration = "Geant4 collaboration",
     title     = "{Geant4: A simulation toolkit}",
     journal   = "Nucl. Instrum. Meth.",
     volume    = "A506",
     year      = "2003",
     pages     = "250",
     doi       = "10.1016/S0168-9002(03)01368-8",
}

@article{Allison:2006ve,
      author         = "Allison, John and Amako, K. and Apostolakis, J. and
                        Araujo, H. and Dubois, P.A. and others",
 collaboration = "Geant4 collaboration",
      title          = "{Geant4 developments and applications}",
      journal        = "IEEE Trans.Nucl.Sci.",
      volume         = "53",
      pages          = "270",
      doi            = "10.1109/TNS.2006.869826",
      year           = "2006",
      reportNumber   = "SLAC-PUB-11870",
}

@Article{Lange:2001uf,
     author    = "Lange, D. J.",
     title     = "{The EvtGen particle decay simulation package}",
     journal   = "Nucl. Instrum. Meth.",
     volume    = "A462",
     year      = "2001",
     pages     = "152-155",
     doi       = "10.1016/S0168-9002(01)00089-4",
}

@article{davidson2015photos,
	author="Davidson, N. and Przedzinski, T. and Was, Z.",
	title = "{PHOTOS interface in C++: Technical and physics documentation}",
      	eprint={1011.0937},
      	archivePrefix={arXiv},
      	primaryClass={hep-ph},
	journal = {Comput. Phys. Commun.},
	volume = {199},
	pages = {86},
	year = {2016},
	doi = {https://doi.org/10.1016/j.cpc.2015.09.013},
}

@PhdThesis{Skwarnicki:1986xj,
  author    = "Skwarnicki, Tomasz",
  title     = "{A study of the radiative cascade transitions between the
                  Upsilon-prime and Upsilon resonances}",
  school = 	 {Institute of Nuclear Physics, Krakow},
  year = 	 {1986},
  note = 	 "{\href{http://inspirehep.net/record/230779/}{DESY-F31-86-02}}",
}

@Article{AdaBoost,
    author = "Freund, Yoav and Schapire, Robert E.",
    title = "A decision-theoretic generalization of on-line learning and an application to boosting",
  journal = 	 "J. Comput. Syst. Sci.",
  volume         = "55",
  pages          = "119",
  year = 	 {1997},
  doi            = "10.1006/jcss.1997.1504",
}

@article{Pivk:2004ty,
      author         = "Pivk, Muriel and Le Diberder, Francois R.",
      title          = "{sPlot: A statistical tool to unfold data distributions}",
      journal        = "Nucl. Instrum. Meth.",
      volume         = "A555",
      pages          = "356-369",
      doi            = "10.1016/j.nima.2005.08.106",
      year           = "2005",
      eprint         = "physics/0402083",
      archivePrefix  = "arXiv",
      primaryClass   = "physics.data-an",
      reportNumber   = "LAL-04-07",
}

@Misc{mciteplus,
  author = 	 {Shell, Michael},
  title = 	 {Mciteplus: Enhanced multicitations},
  howpublished = {\href{http://www.michaelshell.org/tex/mciteplus/} {http://www.michaelshell.org/tex/mciteplus/}},
}

@lhcbreport{LHCb-PUB-2016-021,
      author        = "Anderlini, Lucio and Contu, Andrea and Jones, Christopher
                       Rob and Malde, Sneha Sirirshkumar and Muller, Dominik and
                       Ogilvy, Stephen and Otalora Goicochea, Juan Martin and
                       Pearce, Alex and Polyakov, Ivan and Qian, Wenbin and
                       Sciascia, Barbara and Vazquez Gomez, Ricardo and Zhang,
                       Yanxi",
      title         = "{The PIDCalib package}",
      number        = "LHCb-PUB-2016-021",
      year          = "2016",
      url           = "http://cds.cern.ch/record/2202412",
}

@article{Rogozhnikov:2016bdp,
      author         = "Rogozhnikov, A.",
      title          = "{Reweighting with boosted decision trees}",
      booktitle      = "{Proceedings, 17th International Workshop on Advanced
                        Computing and Analysis Techniques in Physics Research
                        (ACAT 2016): Valparaiso, Chile, January 18-22, 2016}",
      journal        = "J. Phys. Conf. Ser.",
      volume         = "762",
      year           = "2016",
      number         = "1",
      pages          = "012036",
      doi            = "10.1088/1742-6596/762/1/012036",
      eprint         = "1608.05806",
      archivePrefix  = "arXiv",
      primaryClass   = "physics.data-an",
      note           = "\url{https://github.com/arogozhnikov/hep_ml}",
}

@book{Breiman,
  author = 	 {Breiman, L. and Friedman, J. H. and Olshen,
                  R. A. and Stone, C. J.},
  title = 	 {Classification and regression trees},
  publisher = 	 {Wadsworth international group},
  year = 	 {1984},
  address = 	 {Belmont, California, USA},
}

@lhcbreport{Kucharczyk:1756296,
      author        = "Kucharczyk, Marcin and Morawski, Piotr and Witek, Mariusz",
      title         = "{Primary vertex reconstruction at LHCb}",
      institution   = "CERN",
      address       = "Geneva",
      number        = "LHCb-PUB-2014-044",
      month         = "Sep",
      year          = "2014",
      reportNumber  = "LHCb-PUB-2014-044",
      url           = "http://cds.cern.ch/record/1756296",
}

\newpage
% LHCb collaboration author list
% Data extracted on June 24th, 2025 at 5:28pm for paper reference LHCb-PAPER-2025-017
\centerline
{\large\bf LHCb collaboration}
\begin
{flushleft}
\small
R.~Aaij$^{38}$\lhcborcid{0000-0003-0533-1952},
A.S.W.~Abdelmotteleb$^{57}$\lhcborcid{0000-0001-7905-0542},
C.~Abellan~Beteta$^{51}$\lhcborcid{0009-0009-0869-6798},
F.~Abudin{\'e}n$^{57}$\lhcborcid{0000-0002-6737-3528},
T.~Ackernley$^{61}$\lhcborcid{0000-0002-5951-3498},
A. A. ~Adefisoye$^{69}$\lhcborcid{0000-0003-2448-1550},
B.~Adeva$^{47}$\lhcborcid{0000-0001-9756-3712},
M.~Adinolfi$^{55}$\lhcborcid{0000-0002-1326-1264},
P.~Adlarson$^{84}$\lhcborcid{0000-0001-6280-3851},
C.~Agapopoulou$^{14}$\lhcborcid{0000-0002-2368-0147}
% ...

{\footnotesize \it

% ...
$^{14}$Universit{\'e} Paris-Saclay, CNRS/IN2P3, IJCLab, Orsay, France\\
% ...
$^{38}$Nikhef National Institute for Subatomic Physics, Amsterdam, Netherlands\\
% ...
$^{47}$Instituto Galego de F{\'\i}sica de Altas Enerx{\'\i}as (IGFAE), Universidade de Santiago de Compostela, Santiago de Compostela, Spain\\
% ...
$^{51}$Physik-Institut, Universit{\"a}t Z{\"u}rich, Z{\"u}rich, Switzerland\\
% ...
$^{55}$H.H. Wills Physics Laboratory, University of Bristol, Bristol, United Kingdom\\
% ...
$^{57}$Department of Physics, University of Warwick, Coventry, United Kingdom\\
% ...
$^{60}$School of Physics and Astronomy, University of Glasgow, Glasgow, United Kingdom\\
$^{61}$Oliver Lodge Laboratory, University of Liverpool, Liverpool, United Kingdom\\
% ...
$^{69}$Syracuse University, Syracuse, NY, United States\\
% ...
$^{84}$Department of Physics and Astronomy, Uppsala University, Uppsala, Sweden, associated to $^{60}$\\
% ...
}
\end{flushleft}

%\input{~/Downloads/Authorship_LHCb-SD-2024-001-grouped-internal-use-only-do-not-publish-2.tex}
\iffalse
The author list for journal publications is generated from the
Membership Database shortly after 'approval to go to paper' has been
given.  It is available at \url{https://lbfence.cern.ch/membership/authorship}
and will be sent to you by email shortly after a paper number
has been assigned.  
The author list should be included in the draft used for 
first and second circulation, to allow new members of the collaboration to verify
that they have been included correctly. Occasionally a misspelled
name is corrected, or associated institutions become full members.
Therefore an updated author list will be sent to you after the final
EB review of the paper.  In case line numbering doesn't work well
after including the authorlist, try moving the \verb!\bigskip! after
the last author to a separate line.

The authorship for Conference Reports should be ``The LHCb
collaboration'', with a footnote giving the name(s) of the contact
author(s), but without the full list of collaboration names.

The authorship for Figure Reports should be ``The LHCb
collaboration'', with no contact author and without the full list 
of collaboration names.
\fi

\end{document}